\documentclass[12pt]{iopart}
\usepackage{graphicx}
\usepackage[english]{babel}

\renewcommand\vec{\mathbf}
\newcommand\bhat[1]{\mathbf{\hat{#1}}}

\begin{document}

\title[Helitronics for classical and unconventional computing]{Helitronics for classical and unconventional computing}
\author{N. T. Bechler \href{https://orcid.org/0009-0006-6824-3537}{\includegraphics[height=0.75em]{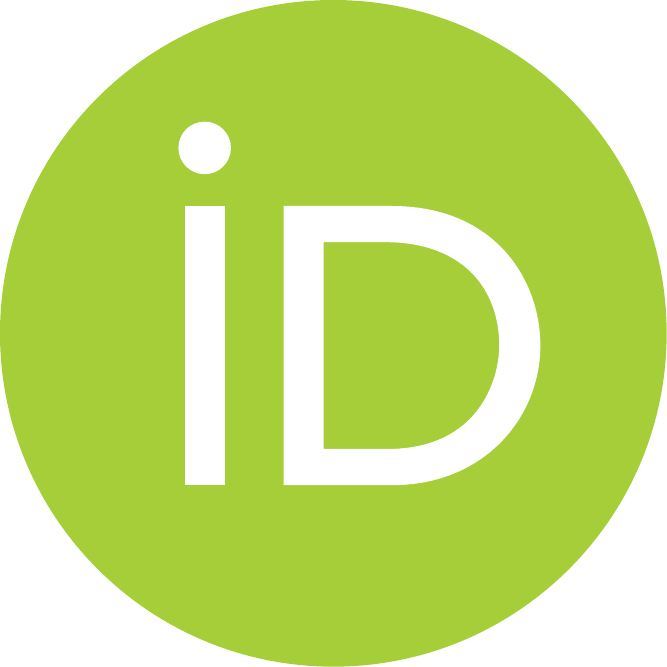}}$^1$ and J. Masell \href{https://orcid.org/0000-0002-9951-4452}{\includegraphics[height=0.75em]{orcid.pdf}}$^{1,2}$}
\address{$^1$ Institute of Theoretical Solid State Physics, Karlsruhe Institute of Technology, 76131 Karlsruhe, Germany}
\address{$^2$ RIKEN Center for Emergent Matter Science (CEMS), Wako 351-0198, Japan}

\begin{abstract}
    Magnetic textures are promising candidates for unconventional computing due to their non-linear dynamics. 
    We propose to investigate the rich variety of seemingly trivial lamellar magnetic phases, e.g., helical, spiral, stripy phase, or other one-dimensional soliton lattices.
    These are the natural stray field-free ground states of almost every magnet.
    The order parameters of these phases may be of potential interest for both classical and unconventional computing, which we refer to as \emph{helitronics}.
    For the particular case of a chiral magnet and its helical phase, we use micromagnetic simulations to demonstrate the working principles of all-electrical (i) classical binary memory cells and (ii) memristor and neuron cells, based on the orientation of the helical stripes.
\end{abstract}

\ead{jan.masell@kit.edu}

\maketitle

\section{Introduction}

A great variety of different systems is currently being investigated for future computation devices and neuromorphic computing will play a key role in the post von-Neumann era.~\cite{Hamdioui2015,Zhu2020}
To build artificial neural systems, memristive building blocks are deemed essential, i.e., electrical components whose resistance is programmable by applying a current. \cite{Yang2013} 
This is due to their ability to store and compute data simultaneously \cite{Zhang2018,Xu2021}, similar to the human brain. 

Magnetic systems have long been used in information technology and they might face a revival by contributing to the emerging field of memristive and neuromorphic devices.
In particular, with the advance of spintronic techniques the electrical manipulation of topological solitons has become possible.
For example, magnetic domain walls can be shifted by electric currents.~\cite{Parkin2009}
Thus, the same current which probes the resistance of a magnetic cell with a domain wall can also shift the domain wall.
If its position determines the resistance, the magnetic cell can serve as a memristor \cite{Wang2009,Thomas2010,Chanthbouala2011,Munchenberger2012} as the relation between the resistance and the pumped charge is linear.~\cite{Sharad2012}
Other solitons like two-dimensional topological skyrmions offer extra degrees of freedom in their translational dynamics.
They have thus been suggested as alternatives for one-dimensional domain walls in many systems, including shift-register memories \cite{Fert2013,Mueller2017} or artificial synapses~\cite{Li2017,Prychynenko2018}.
Other studies also suggest to fully exploit the two-dimensional mobility of skyrmions and related textures, such as dense fabrics \cite{Bourianoff2018} or individual exotic biskyrmions~\cite{Ribeiro_de_Assis2023}.

\begin{figure}
	\center
	\includegraphics[width=15cm]{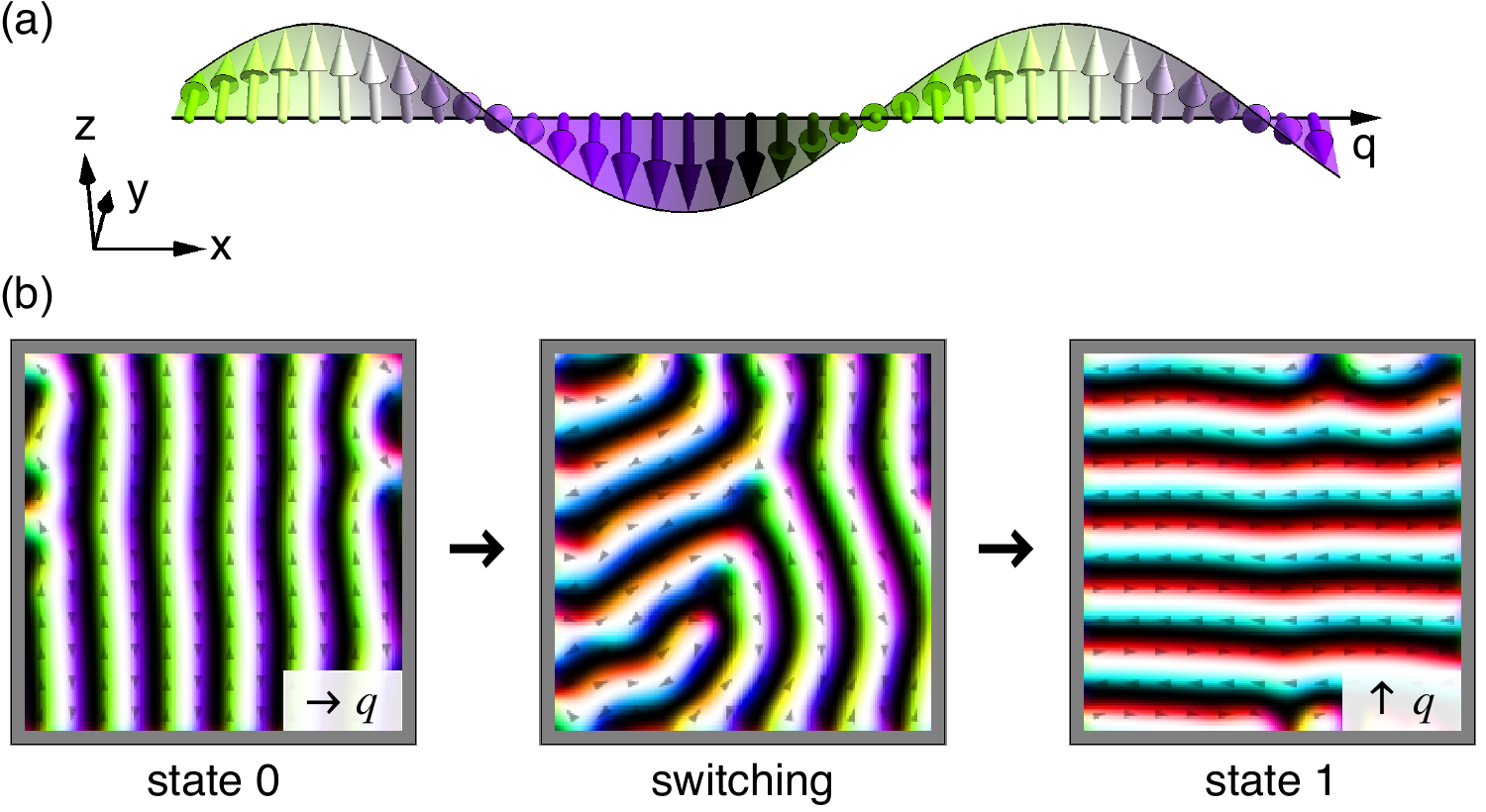}
	\caption{
        Helical order and switching of the orientation of the $\bhat{q}$-vector.
        (a) Schematic magnetization texture in the helical phase.
        (b) Example of two distinct helitronic states $0$ ($\bhat{q}\parallel\bhat{x}$) and $1$ ($\bhat{q}\perp\bhat{x}$) in a finite size system.
        Additionally, the transient state during switching is shown. 
        Result of a simulation with spin transfer-torque, see \sref{sec2.3}.
        Color indicates the direction of the magnetization.
        Small gray arrows indicate the in-plane component of the magnetization.
        The surrounding gray background is vacuum.
        }
	\label{fig1:schematic_render}
\end{figure}

However, isolated domain walls or skyrmions are usually artificially created metastable states.
The thermodynamic ground state of almost every magnet in the absence of external magnetic fields is a demagnetized stripe/spiral/helical phase, see \fref{fig1:schematic_render}(a).
These states which can be viewed as a labyrinth of one-dimensional topological solitons, are coming back into the focus of spintronic research due to their unique symmetry.
They have been shown to work as emergent inductors~\cite{Nagaosa2019,Yokouchi2020,Kitaori2021}, lead to non-reciprocal charge and spin transport~\cite{Rikken2001,Jiang2020,Ishizuka2020,Okumura2019}, or may actively pump spin and charge as Archimedean screws~\cite{Tserkovnyak2005,delSer2021,Kurebayashi2022}.
Moreover, they can serve as a non-trivial background for other topological excitations such as skyrmions~\cite{Muller2017,Knapman2021,Song2022} or dislocations~\cite{Dussaux2016,Schoenherr2018,Stepanova2022,Azhar2022}.
Only recently, it has been shown that the orientation vector $\vec{q}$ of this lamellar phase can be combed and steered by external magnetic fields \cite{Bauer2017,Liu2019} and even electric currents~\cite{Masell2020}. 

In this work, we propose to exploit the intrinsic properties of the demagnetized state of magnets for spintronic applications.
We focus on two-dimensional chiral magnets whose ground state is the helix with a well-defined ordering vector $\vec{q}$, see \fref{fig1:schematic_render}(a).
Based on the helical orientation $\vec{q}$, we can define logical states and design \emph{helitronic} devices.
The model for a chiral magnet is introduced in \sref{sec2} where we define an order parameter, discuss electrical read-out based on anisotropic magneto-resistance (AMR), and show simulation results for combing the helical orientation $\vec{q}$ in small square-shaped cells.
We discuss examples for binary memory cells (HRAM) in \sref{sec3} and non-binary memory cells with memristive or neuromorphic properties in \sref{sec4}.
We show that helitronic states may be electrically controlled and measured via their characteristic resistance and that they hold great potential for classical and unconventional computing devices.

\section{The helical orientation as a new order parameter}
\label{sec2}

Chiral magnets, i.e., magnets with broken inversion symmetry in the unit cell are a particularly well-studied system which exhibits lamellar magnetic order.
The crystalline chirality in combination with spin-orbit coupling (SOC) gives arise to an antisymmetric exchange interaction known as Dzyaloshinskii-Moriya interaction (DMI).~\cite{Dzyaloshinsky1958, Moriya1960}
The competition of DMI and magnetic stiffness is known to stabilize long-ranged \emph{helical} order, see \fref{fig1:schematic_render}(a), where the magnetization rotates in the plane perpendicular to the $\bhat q$-vector.~\cite{Bak1980,Uchida2006}
Such systems have recently regained massive attention due to the stabilization of topologically non-trivial skyrmions at finite temperature and externally applied magnetic field.~\cite{Muehlbauer2009, Yu2010}
However, the thermodynamic ground state without external field is a multidomain helical phase.~\cite{Schoenherr2018}
We will therefore use two-dimensional chiral magnets as a conceptual model for helitronic applications throughout this study.

\subsection{The model: two-dimensional chiral magnet}
\label{sec2.1}

To ensure a maximum of universality, we consider the most simple model for two-dimensional chiral magnets.
It comprises only the magnetic stiffness $A$ and DMI $D$,
\begin{eqnarray}
    E[\bhat{m}] = \int A \, \left( \vec{\nabla} \bhat{m} \right)^2 + D \, \bhat{m} \cdot \left( \vec{\nabla} \times \bhat{m} \right) \, d^2 r
	\label{eq1:free_energy_functional}
\end{eqnarray}
where $\bhat{m} = \vec{M}/M_s$ is the normalized magnetization and $M_s$ the saturation magnetization.
All other contributions to the energy are disregarded for the sake of simplicity as they will only quantitatively alter the results.
In particular, anisotropic interactions which affect the direction of the $\bhat{q}$-vector are also neglected.
Such anisotropies are higher order in SOC and, thus, usually small compared to the exchange and DMI.
In particular, they are small compared to the torque exerted by the applied switching currents, see \sref{sec2.3}. 
Moreover, we also neglect dipolar interactions as the net magnetic moment vanishes in the helical phase which strongly suppresses the effect of demagnetization.~\cite{Dussaux2016}

For all simulations and analytical calculations, we use the established material parameters of FeGe \cite{Beg2015}, i.e., $A = 8.75 \, \mathrm{pJ/m}$, $D = 1.58 \, \mathrm{mJ/m^2}$, and $M_s = 384 \, \mathrm{kA/m}$.
The resulting wavelength of the helical order is $\lambda=69.59 \, \mathrm{nm}$.
However, the results presented in the following are sufficiently generic to be applicable also to systems with much smaller wavelength and, thus lateral extension of the devices.

\subsection{The helical orientation order parameter $h_y$}
\label{sec2.2}

For an infinitely large system, the energy functional \eref{eq1:free_energy_functional} is minimized by a monodomain helical texture 
\begin{eqnarray}
	\bhat{m}(\vec{r}) = (\bhat{z} \times \bhat{q})\cos(\vec{q}\cdot \vec{r} + \varphi) +\bhat{z}\sin({\vec{q}\cdot\vec{r}} + \varphi)
	\label{eq2:magnetization}
\end{eqnarray} 
with $\vec{q} = D/(2  A) \, \bhat{q}$ the uniform orientation vector of the helix and $\varphi$ an arbitrary phase.
Since the different orientations in the plane are energetically degenerate, one usually observes a multidomain state with patches of different helical orientations $\bhat{q}$.
Moreover, in finite size systems the $\bhat{q}$-vector is weakly pinned by shape anisotropy.

\begin{figure}
	\center
	\includegraphics[width=15cm,keepaspectratio]{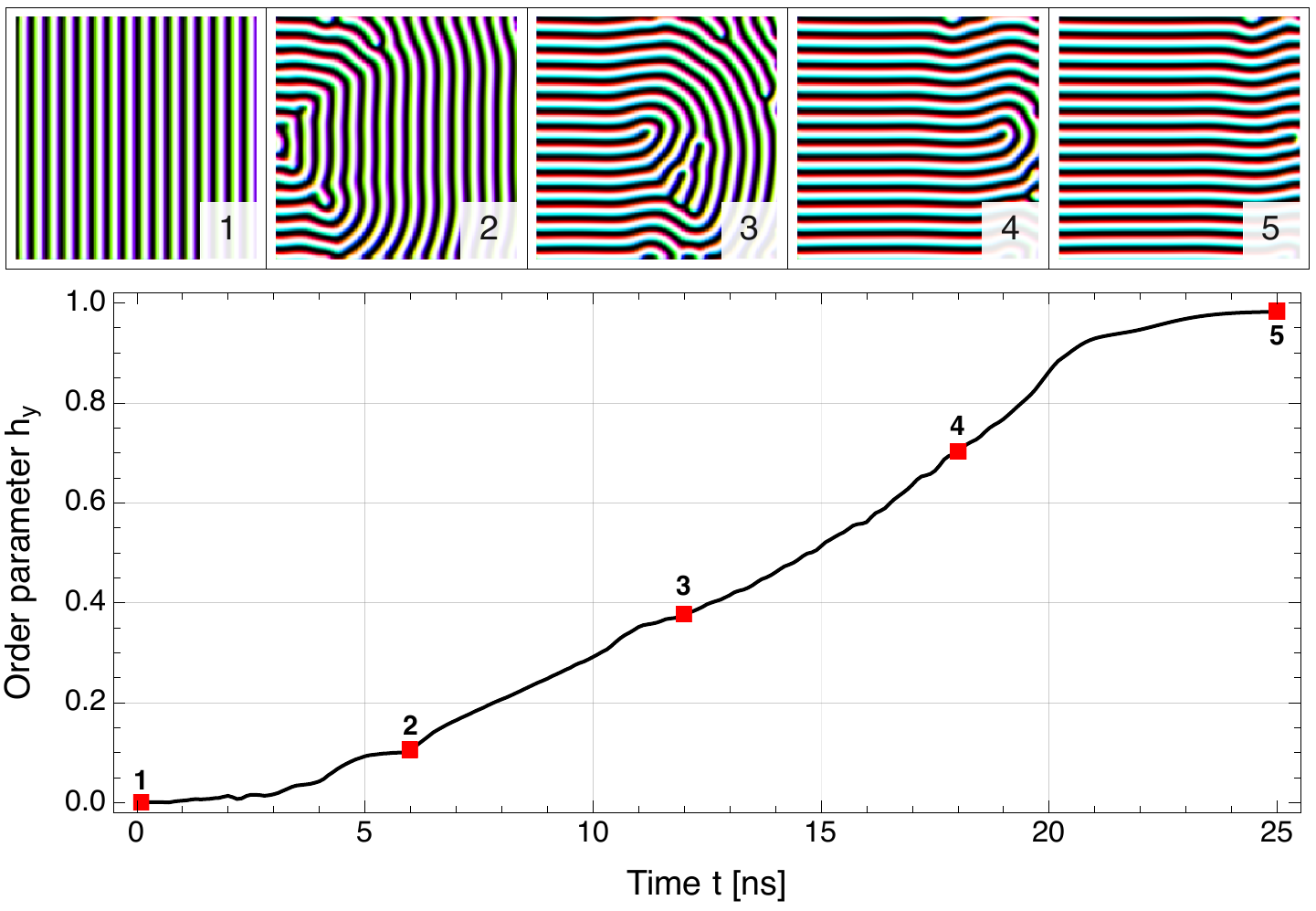}
	\caption{
        Detailed switching process of the helical orientation by spin-transfer torque in a square-shaped cell.
        Upper panels show the realspace magnetization during the switch.
        Lower plot shows the evolution of the order parameter $h_y$ as function of time.
        Numbers indicate the position of the realspace panels. 
        The initial texture (panel 1) with $\bhat{q}=\bhat{x}$ and $h_y=0$ is pushed out to the right and replaced with a phase with $\bhat{q}=\bhat{y}$ and $h_y=1$ (panel 5).
        The system size is $16\lambda \times 16\lambda$.
        A constant current $\vec{j} = -32 \cdot 10^{10}\, \mathrm{A/m^2} \,\bhat{x}$ is applied homogeneously, see \sref{sec2.3} for details. 
        }
	\label{fig2:detailed_switch}
\end{figure}

In order to quantify the progress of the switching and reorientation of the helix, a quantitative order parameter is needed.
While tracking the direction of the $\bhat{q}$-vector seems natural, it is not well defined in the case of multidomain states which naturally appear during switching as in \fref{fig1:schematic_render}(b).
We therefore make use of the Fourier amplitude $m^z_{\vec{q}}$ of the z-component of the magnetization 
\begin{eqnarray}
	m^z_{\vec{q}} = \frac{1}{V} \int \bhat{m}^z(\vec{r}) \, \e^{-i\mkern1mu \, \vec{q}\cdot\vec{r}} dV\,\,.
	\label{eq3}
\end{eqnarray} 
Throughout this study, we focus on the particular orientation $\vec{q}=\frac{2\pi}{\lambda}\bhat{y}$ and accordingly define the helical orientation order parameter $h_y$ as
\begin{eqnarray}
	h_y = \left|\,2\, m^z_{\vec{q}}\,\right|_{\vec{q}=\frac{2\pi}{\lambda}\bhat{y}} = 
    \left\{
    \begin{array}{lr}
        0 & : \mathrm{state\ 0}\\
        1 & : \mathrm{state\ 1}
    \end{array}
    \right.
	\label{eq4}
\end{eqnarray}
where the states $0$ and $1$ refer to \fref{fig1:schematic_render}(b) and the factor $2$ was added to account for the symmetry in $\vec{q}$.
The relation between the realspace magnetization and the order parameter $h_y$, which corresponds to the volume fraction in a state with $\vec{q}=\frac{2\pi}{\lambda}\bhat{y}$, is illustrated in \fref{fig2:detailed_switch} during a switching process.
The details of this process are discussed in \sref{sec2.3}.

\subsection{Magnetization dynamics and spin-transfer torque}
\label{sec2.3}

To describe the magnetization dynamics during the switching process, we employ the Landau-Lifshitz-Gilbert (LLG) equation~\cite{Gilbert2004}
\begin{eqnarray}
	\frac{d\bhat{m}}{dt} = - \gamma \, \bhat{m} \times \vec{B}_{\mathrm{eff}} + \alpha \, \bhat{m} \times \frac{d\bhat{m}}{dt} + \tau
 \label{eq5}
\end{eqnarray}
with $\gamma$ the gyromagnetic ratio, $\vec{B}_{\mathrm{eff}}=-M_s^{-1}\,\delta E/\delta\bhat{m}$ the effective magnetic field and $\alpha$ the Gilbert damping parameter.
Additional torques which lead to a reorientation are included via $\tau$, for example magnetic fields~\cite{Bauer2017,Liu2019}, electric currents~\cite{Hals2019,Masell2020}, or temperature gradients~\cite{Wang2016}.
For all-electrical switching, following Ref.~\cite{Masell2020}, we include Zhang-Li spin-transfer torque (STT)~\cite{Zhang2004} which takes the form  
\begin{eqnarray}
	\tau_\mathrm{STT} = \frac{P \, \mu_B}{e \, M_s (1+\beta^2)} \left[ (\vec{j}\cdot \vec{\nabla}) \, \bhat{m} - \beta \,\bhat{m} \times (\vec{j}\cdot \vec{\nabla}) \, \bhat{m} \right]
    \label{eq6}
\end{eqnarray}
with $P$ the spin polarization, $\mu_B$ is the Bohr magneton, $e>0$ the electron charge, $\beta$ the non-adiabatic damping parameter and $\vec{j}$ the current density.
This torque stems from the misalignment of the conduction electron spin and the local magnetization.
Accordingly, it is maximized for $\vec{j}\parallel\vec{q}$ and vanishes for $\vec{j} \perp \vec{q}$ where the magnetization is translation invariant in the direction of electron flow, which makes this $\bhat{q}$-vector orientation a fixed point of the equation of motion.
We solve equation~\eref{eq5} numerically using a modified version of the GPU-accelerated micromagnetic solver MuMax3.~\cite{Vansteenkiste2014}
In the modified version, derivatives in the effective magnetic field $\vec{B}_\mathrm{eff}$ are discretized by fourth order finite differences, similar to the code used in Ref.~\cite{Yu2020}, which improves accuracy and minimizes numerical artifacts such as lattice anisotropies.
Throughout this paper we use $\alpha = 0.1$, $\beta = 0.2$, and $P=1$.
The system size is always chosen in integer multiples of the helical wavelength $\lambda$ and indicated in the respective sections.
For technical reasons, the numerical lattice is a monolayer of three-dimensional cuboids with discretization $a_x=a_y=\lambda/16$ along the x- and y-direction and $a_z=3\,\mathrm{nm}$ along the z-direction.

\subsection{Electrical switching of the orientation vector $\hat{q}$}
\label{sec2.4}

When applying a current to a finite size system, we observe that the initial state is pushed out and replaced by a newly incoming helical state.
An example of such a transition is shown in \fref{fig2:detailed_switch} where the initial state is a helix with $\bhat{q}\parallel\bhat{x}$ and, thus, $h_y=0$.
The incoming state when applying a current of $\vec{j} = -32 \cdot 10^{10}\, \mathrm{A/m^2} \,\bhat{x}$ is translation invariant in the direction of the current, i.e. $\bhat{q}\perp\vec{j}$ and $h_y=1$, as previously reported in Ref.~\cite{Masell2020}.
Notably, the order parameter $h_y$ varies linearly with time, $h_y\propto t$.
This linear behavior is only corrupted near the polarized cases $h_y=0$ or $1$, where the topologically non-trivial interface between the helical state with distinct orientations has to overcome an energy barrier to enter/exit at the edges of the sample.

\begin{figure}
	\center
	\includegraphics[width=15cm]{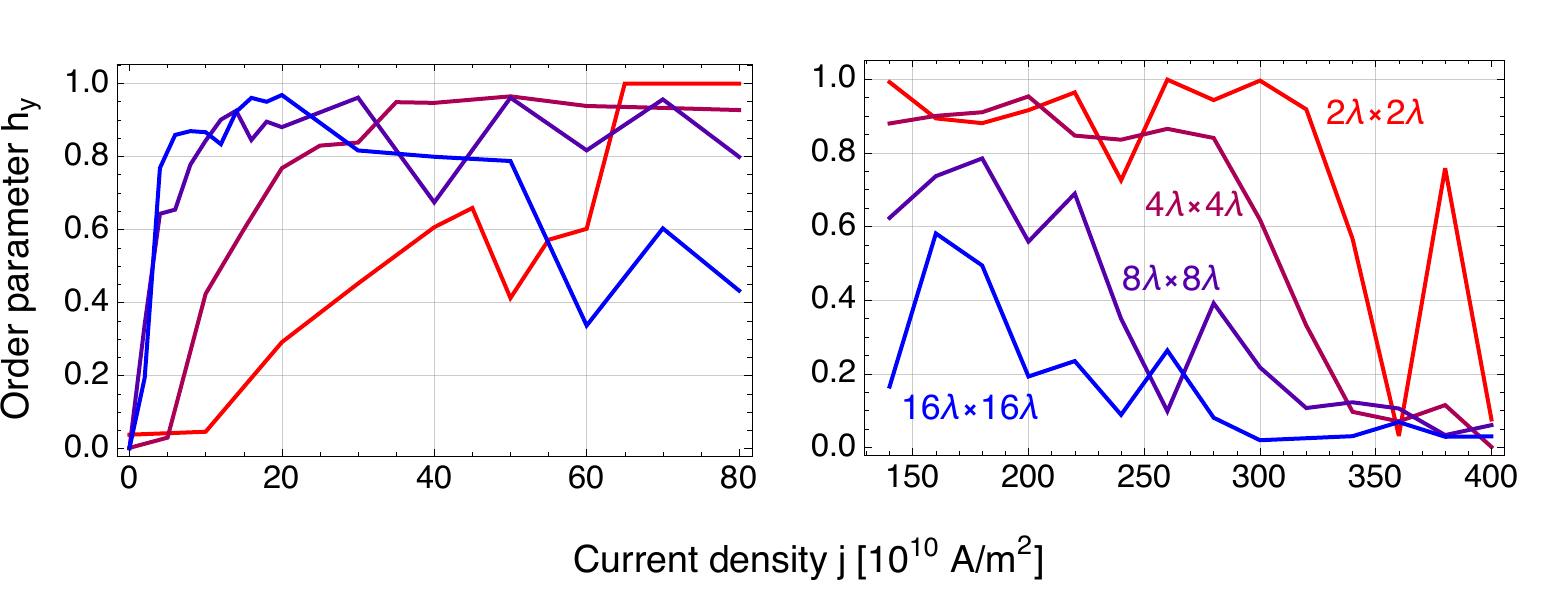}
	\caption{
        Helical orientation order parameter $h_y$ in the long-time limit as function of current density.
        Low current densities from depinning transition to saturation are shown in the left panel.
        High current densities are shown in the right panel, extending beyond the stability regime.
        Color encodes the size of the square-shaped system as indicated.
        The current is applied in $-\bhat{x}$-direction. 
        Results are obtained by running simulations until a steady state or characteristic dynamical state is reached. 
        The noise stems from the high sensitivity of the order parameter with respect to defects, in particular in the high current limit where no steady state exists.
        }
	\label{fig3:helitronics_showcase}
\end{figure}

Consequently, since an energy barrier has to be overcome to introduce the new helical state, a lower critical current density is expected.
Moreover, an upper critical current density was calculated in Ref.~\cite{Masell2020} for which the magnetization gets polarized along the current direction.
For $\vec{q}\parallel\vec{j}$ it is given by 
\begin{eqnarray}
	j_\mathrm{c}^\mathrm{pol} = \frac{2\alpha (1+\beta)^2 \, \gamma \, e}{\left| \alpha - \beta \right| \mu_\mathrm{B}} (D-A \, q)
    \approx 6.92\cdot10^{12} \, \mathrm{A/m^2}
    \label{eqSomething}
\end{eqnarray}
for FeGe and the equilibrium $q=D/(2A)$.
\Fref{fig3:helitronics_showcase} shows the order parameter $h_y$ as a function of system size and current density in the steady state, i.e., in the limit of long times.
The data is noisy as the order parameter is highly sensitive to small deviations from perfectly polarized helical order.
Nevertheless, a clear dependence on the applied current and system size is observed.
For the smallest system sizes, $2\lambda\times2\lambda$, we find (i) a lower critical current density $|j_{\mathrm{c},1}| \approx 10 \cdot 10^{10} \, \mathrm{A/m^2}$ for the onset of switching, (ii) an intermediate critical current density $|j_{\mathrm{c},2}| \approx 65 \cdot 10^{10} \, \mathrm{A/m^2}$ for full switching, and (iii) an upper critical current density $|j_{\mathrm{c},3}| \approx 320 \cdot 10^{10} \, \mathrm{A/m^2}$ which is the maximal current density for switching above which the magnetization does not converge to a steady state anymore.
For larger system sizes, these critical current densities are lowered such that $|j| \approx 10 \cdot 10^{10} \, \mathrm{A/m^2}$ is sufficient for full switching in systems larger than $8\lambda\times8\lambda$.

This behavior can be qualitatively understood as the only sources for pinning are the edges and corners of the system which become increasingly irrelevant as the system size increases.
However, this picture only holds as long as the force density is homogeneous in the sample, i.e., it breaks down once a relevant fraction of the system is switched such that $\vec{q}\perp\vec{j}$ and, thus, $(\vec{j}\cdot\nabla)\bhat{m}=0$ in this area.
Consequently, the critical current for full switching should become independent of the system size for large systems, which we cannot confirm with our simulations, probably because the system sizes considered here are still quite small.
Note, also, that smaller current densities and larger system size imply larger switching times $T \sim L/j$.
This sets limits to both the technological relevance and the minimal currents that we can numerically investigate with reasonable effort.
Moreover, the upper critical current density is lowered as the system size increases.
This is a consequence of defects entering from the transverse edges as well as multiple instabilities of the helical phase, c.f. Ref.~\cite{Masell2020}, which are suppressed in small systems.

\subsection{Electrical read-out via anisotropic magneto-resistance}
\label{sec2.5}

Operating a helitronics device requires a mechanism to probe the helical orientation.
The different lamellar orientations are reminiscent of the texture in the liquid crystals in LCD displays and optical read-out would certainly be elegant.
However, for practical integration into CMOS technology a purely electrical read-out mechanism is preferred.
We therefore consider a simple model for AMR, i.e., the dependence of the resistivity on the orientation of the magnetization, as explained in the following.
The state of the helitronic cell can then be read out electrically by measuring the resistance of the cell.
The mechanism behind the AMR-effect is anisotropic scattering of conduction electrons due to SOC. \cite{Smit1951}
The resistance depends on the angle between the magnetization $\bhat{m}$ and the current density $\vec{j}$ \cite{Doering1938,McGuire1975}
\begin{eqnarray}
	\rho = \rho_{\perp} +(\rho_{||}-\rho_{\perp}) \, \cos^2\angle(\bhat{m},\vec{j}) \,\,.
	\label{eq7}
\end{eqnarray}
$\rho_\perp$ and $\rho_\parallel$ denote the distinct resistances perpendicular and parallel to the magnetization $\bhat{m}$.
As this model describes only uniformly magnetized samples, additional effects need to be considered when discussing non-collinear magnets.
For example, sizable changes in the resistance are expected if $\vec{q}$ is large, which results in a strong spin-transfer torque and, by reciprocity, a strong $\bhat{q}$-dependent resistance.~\cite{Rikken2001,Jiang2020,Ishizuka2020,Okumura2019,Aqeel2020}

For the purpose of this study, however, we follow the approaches in Refs.~\cite{Prychynenko2018,Bohlens2010, Krueger2011}.
We consider the continuity equation and Ohm's law,
\begin{eqnarray}
	\frac{\partial \rho_\mathrm{e}}{\partial t} + \vec{\nabla}\vec{j} = 0 
    \quad\mathrm{and}\quad
    \vec{j} = - \underline{\sigma} \vec \nabla \phi \,\,,
	\label{eq8}
\end{eqnarray}
which relates the charge density $\rho_\mathrm{e}(\vec{r})$, current density $\vec{j}(\vec{r})$, electrostatic potential $\phi(\vec{r})$, and conductivity tensor $\underline{\sigma}(\vec{r})$.
For homogeneous charge densities, $\frac{\partial \rho_\mathrm{e}}{\partial t}=0$, we obtain the two equivalent differential equations
\begin{eqnarray}
	\vec{\nabla}(\underline{\sigma}\vec{\nabla}\phi) = 0
    \quad\mathrm{or}\quad
    \int_{\partial S}\underline{\sigma}\vec{\nabla}\phi \, d^2r = 0 \,\,.
	\label{eq9}
\end{eqnarray}
Together with the boundary conditions $\phi(x=0)=U$ on the left edge and $\phi(x=L_x)=0$ on the right edge of the sample, which defines the voltage $U$, equation~\eref{eq9} uniquely determines the potential $\phi(\vec{r})$.
The conductivity tensor $\underline{\sigma} =\underline{\rho}^{-1}$ follows from equation~\eref{eq7} in two dimensions as
\begin{eqnarray}
\underline{\sigma}(\bhat{m}) = \frac{1}{\rho_0}\left[ 1\!\!1 - \frac{3 a}{3-a}
	\left(
    \matrix{ \chi(m_z)\, m_x^2 - \frac{1}{3} & \chi(m_z)\, m_x m_y \cr
	\chi(m_z)\, m_x m_y & \chi(m_z)\, m_y^2 - \frac{1}{3} \cr} \right) \right]
	\label{eq10}
\end{eqnarray}
with $\chi(m_z) = 1/[1 + a(\frac{2}{3} - m_z^2)]$ introduced for better readability, $\rho_0 = \frac{1}{3}(\rho_\parallel+ 2 \rho_{\perp})$ the mean resistivity of a demagnetized sample, and $a = (\rho_\parallel - \rho_\perp)/\rho_0$ the AMR-ratio.
Note that in two dimensions $\rho$ has the same units as $R$.
In the following, we consider $a = 10\%$.
For computing the time-dependent resistance $R(t)$ of the device during a switching process, we first run a simulation of the magnetization dynamics with a constant and homogeneous current density $\vec{j}$.
We checked that a spatially homogeneous current density is a good approximation for $a = 10\%$.
From the time-dependent magnetization data we later compute the resistance $R(t)=U(t)/I$ by numerically solving equations~\eref{eq9} and~\eref{eq10} on the numerical lattice.
We finally normalize the resistance $R(t)$ by 
\begin{eqnarray}
    R_0 = \frac{\rho_0}{3} \sqrt{9+3a-2a^2} \approx 1.0154 \, \rho_0
	\label{eq11}
\end{eqnarray}
which is the analytical result for the resistance of a fully switched cell of size $n\lambda \times n\lambda$ with $h_y=1$, $\varphi = 0$ and $a = 10\%$.
In turn, the resistance of a perpendicularly oriented helical phase with $\bhat{q}\parallel\bhat{x}$ is  $R_1= \rho_0 \frac{3 - a}{3} \approx 0.967 \rho_0$.
Putting these together, we can estimate the maximal resistance change $\delta R = 1 - R_1/R_0.$
Consequently, for an AMR of $a = 10\%$ the change in resistance between two helitronic states will be less than $\delta R\approx 8.3\%$ which can be optimized by the choice of materials or exploiting other effects than this elementary form of AMR, e.g., in Weyl semimetals.

\section{Binary cell: HRAM}
\label{sec3}

\begin{figure}
	\center
	\includegraphics[width=15cm,keepaspectratio]{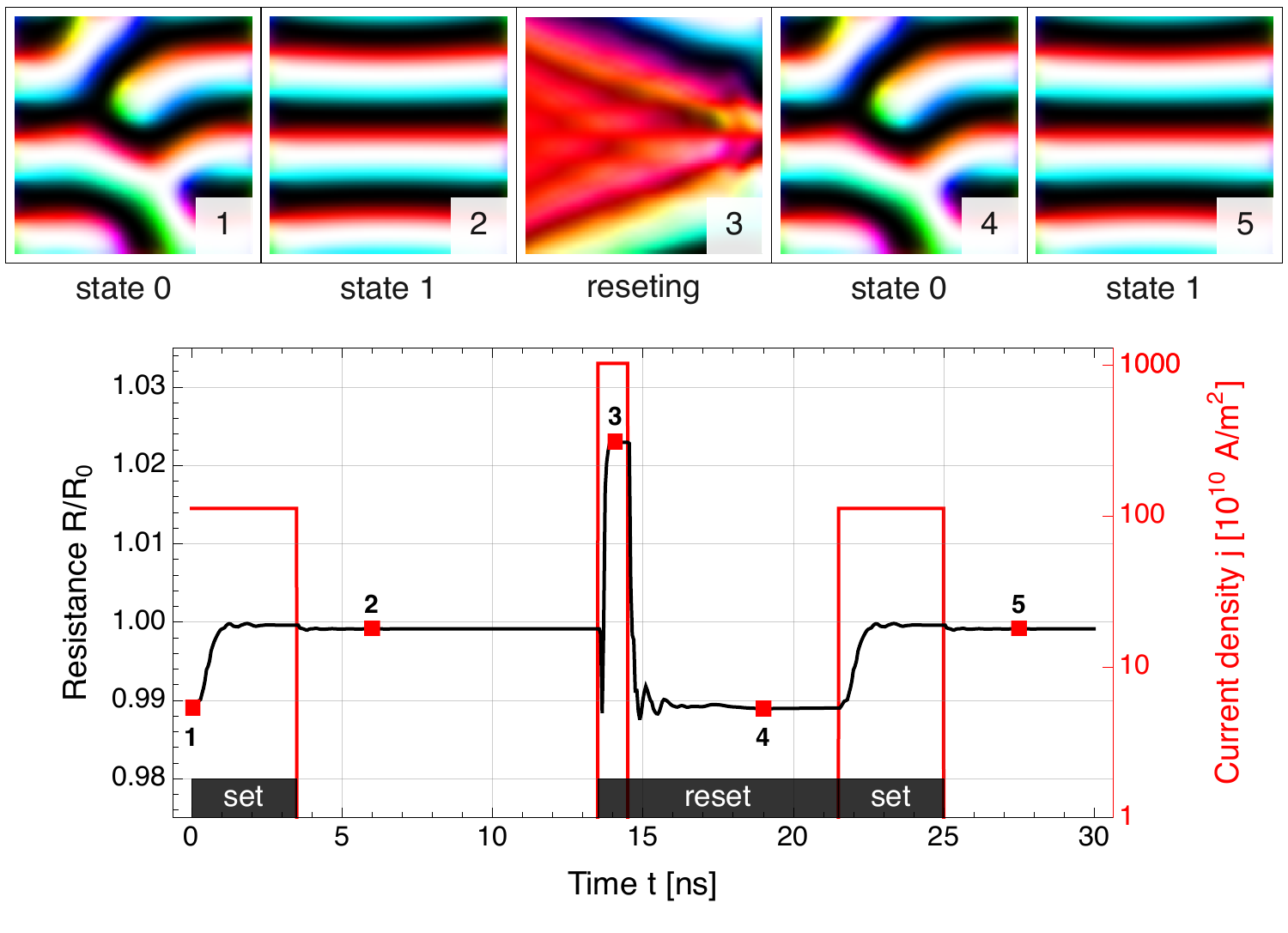}
	\caption{
        Simulation of a helitronic binary memory cell (HRAM).
        The upper panels show the magnetization during a series of operations: (1) in state $0$,  (2) in state $1$ after switching with $\vec{j} = -112 \cdot 10^{10}\,\mathrm{A/m^2} \; \bhat{x}$, (3) while resetting to state $0$ with $\vec{j} = -1024 \cdot 10^{10}\,\mathrm{A/m^2} \; \bhat{x}$, (4) in state $0$ after completing the reset, and (5) in state $1$ after switching again.
        The lower plot shows the rescaled resistance $R/R_0$ (black line) as function of time for the same simulation.
        The position of the above panels is indicated by red squares and labeled accordingly.
        The red curve indicates the current density.
        The system size is $3\lambda\times3\lambda$.
        }
	\label{fig4:MRAM_cell}
\end{figure}

A device which can write and read two distinct helical states can in principle be operated as a helitronic bit (HRAM).
For example, in a four-terminal device we could apply currents along the $\bhat{x}$- or $\bhat{y}$-direction and thereby polarize the $\bhat{q}$-vector perpendicular to the respective current directions.
For reading the states only two terminals are required as the states can be distinguished according to their resistance $R$ for a fixed probing direction.

However, a two-terminal device is easier to manufacture and operate and, thus, more desirable.
Simulation results for a HRAM cell based on the model in \sref{sec2} are shown in \fref{fig4:MRAM_cell}.
The size of the device is only $3\lambda\times3\lambda$ which is sufficient to host the two elementary states which form the binary basis of our device.
In contrast to the above introduced four-terminal device with two complementary ordered states, we now distinguish between a disordered state $0$ (panel 1) and an ordered state $1$ (panel 2).
The states can be confidently identified by their respective resistance, see the black line in the graph in \fref{fig4:MRAM_cell}.
The current densities required for writing the states $0$ and $1$ can be estimated from the steady state analysis in \fref{fig3:helitronics_showcase}. 
To write the ordered state $1$, we apply a current density $\vec{j} = -112 \cdot 10^{10}\,\mathrm{A/m^2} \; \bhat{x}$ for $3.5 \,\mathrm{ns}$ which is large enough to overcome the pinning barriers for full switching.
Once written, the state $1$ is metastable and protected by an energy barrier.
To write the disordered state $0$, we apply a current density $\vec{j} = -1024 \cdot 10^{10}\,\mathrm{A/m^2} \; \bhat{x}$ for $1\,\mathrm{ns}$ which is above the Walker breakdown of the helical phase~\cite{Masell2020} and polarizes the magnetization $\bhat{m}\parallel-\vec{j}$, see panel (3). 
Once this high current is switched off, the magnetization freezes in a local energy minimum which is a disordered helical state, see panel (4).
As the system size is chosen sufficiently small, however, this disordered state appears to be unique and is reproducible with high fidelity.
Finally, see panel (5), the cell can be switched to state $1$ again by the smaller writing current.

While a two-terminal device might have its benefits, the drawbacks of this particular implementation are also obvious.
Despite an AMR-ratio  of $a = 10\%$, the resistances of the ordered and disordered state differ by only 1\%.
This is far below the 8.3\% hypothetically achievable difference in a device that uses two perpendicular orientations of the $\bhat{q}$-vector, see \sref{sec2.5}.
Moreover, the disordered state might not be unique or the ordered state might have defects due to material impurities or temperature fluctuations, which will further reduce the fidelity for reading the state.
Finally, shape anisotropy and the effects of thermal fluctuations will be most pronounced for the smallest system sizes which implies that small HRAM cells are also prone to quickly fading memory.
These issues could be overcome, for example, by considering materials with shorter wavelengths and much larger directional anisotropies which put larger bounds on the energy landscape.

\section{Non-binary cell: Helitronic memristor and artificial neuron}
\label{sec4}

In contrast to binary memory cells, a memristive cell needs a continuous spectrum of states.
A helitronics cell is therefore a natural candidate to consider as the helical orientation is a continuous parameter: 
In sufficiently isotropic systems, ideally disks, the $\bhat{q}$-vector can point in any direction and can be smoothly rotated.
Moreover, a memristor needs to \emph{memorizes} the number of applied current pulses by altering the resistance by a constant shift for every pulse and, of course, a mechanism to reset the memory.
If the memory is continuously fading with time, the memristor can even serve as an artificial neuron.
In this section, we show that a helitronic system can serve as a memristor or artificial neuron by again considering a simple square-shaped geometry for the magnetic sample.

\begin{figure}
	\center
	\includegraphics[width=15cm,keepaspectratio]{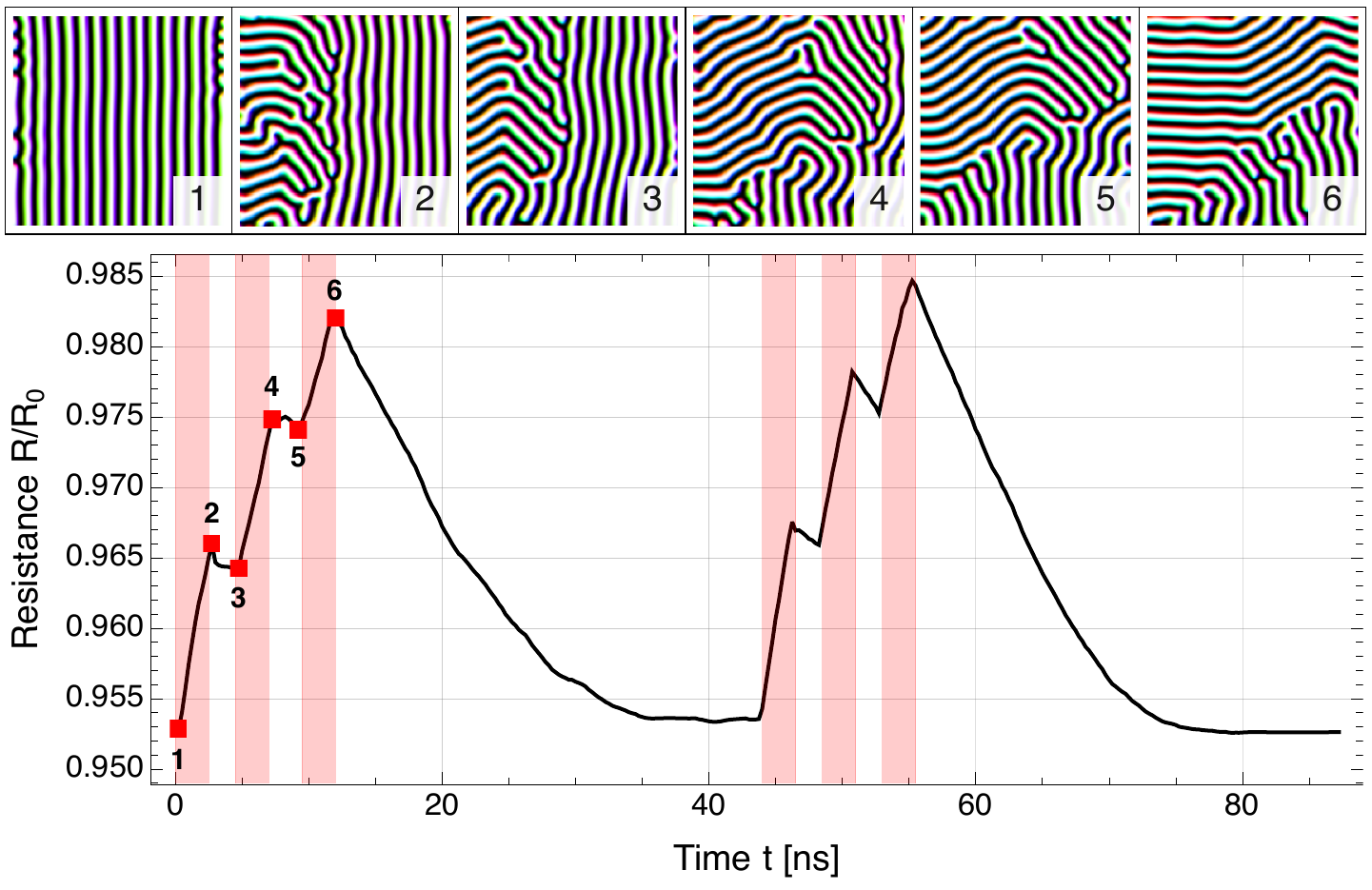}
	\caption{
        Simulation of a helitronic memristor.
        The upper panels show the magnetization during operation at different times indicated by red squares in the graph.
        The graph shows the normalized resistance $R/R_0$ as function of time (black line) and indicates where write currents $\vec{j}_\mathrm{w} = -64 \cdot 10^{10}\,\mathrm{A/m^2} \; \bhat{x}$ are applied (red shaded areas).
        A constant reset current $\vec{j}_\mathrm{r} = - 24 \cdot 10^{10}\,\mathrm{A/m^2} \; \bhat{y}$ is continuously applied for the whole operation time.
        The system size is $16 \lambda \times 16 \lambda$.
        } 
	\label{fig5:memristor}
\end{figure}

Unlike in \sref{sec3}, we now consider a significantly larger system and a four-terminal architecture which allows to apply current pulses in every direction, in particular along the $\bhat{x}$- and $\bhat{y}$-direction independently.
The two orthogonal directions for current pulses are used to imprint the two orthogonal helical orientations, $\vec{q}\parallel\bhat{x}$ and $\vec{q}\perp\bhat{x}$.
Simulation results for the operation of a helitronic memristor with artificially fading memory are shown in \fref{fig5:memristor}.
The system size is $16 \lambda \times 16 \lambda$ and the normal state of the device is an almost perfectly ordered helical phase with $\vec{q}\parallel\bhat{x}$, see panel (1).
Three $2.5 \, \mathrm{ns}$ current pulses of $\vec{j}_\mathrm{w} = -64 \cdot 10^{10}\,\mathrm{A/m^2} \; \bhat{x}$ are applied every $2 \, \mathrm{ns}$.
With every pulse, a certain area of the initial order is pushed out of the system over the right edge as the orthogonal state with $\vec{q}\perp\bhat{x}$ penetrates from the left edge.
Besides small corrections from the interface between the two regions, the change in resistance is linear in the switched area which, itself, is linear in the duration and amplitude of the current pulse.
With sufficient pinning, which is intrinsic to the helical phase, the state is non-volatile and the device can therefore serve as a memristor.

In \fref{fig5:memristor}, a constant reset current with $\vec{j}_\mathrm{r} = - 24 \cdot 10^{10}\,\mathrm{A/m^2} \; \bhat{y} \perp \vec{j}_\mathrm{r}$ is applied.
The purpose of the reset current $\vec{j}_\mathrm{r}$ is to artificially erase the memory of the cell and, continuously, reset it to the normal state $\vec{q}\parallel\bhat{x}$ which enters the system from the bottom edge.
As can be seen in panels (2-6), the combined action of the continuous reset current $\vec{j}_\mathrm{r}$ and pulses of $\vec{j}_\mathrm{w}$ creates a multidomain helical state.
The graph shows the normalized resistance $R/R_0$ during the operation which essentially is a linear superposition of the volume fractions of the two orthogonal phases.
Note that the quantization of current pulses here is not perfect due to the fading memory.
After the three pulses, no switching current is applied for $30 \, \mathrm{ns}$ letting the cell reset back to its normal state. 
Afterwards, the process is repeated and it is confirmed that the characteristic resistance curves are reproducible.

\begin{figure}%
	\center
	\includegraphics[width=15cm,keepaspectratio]{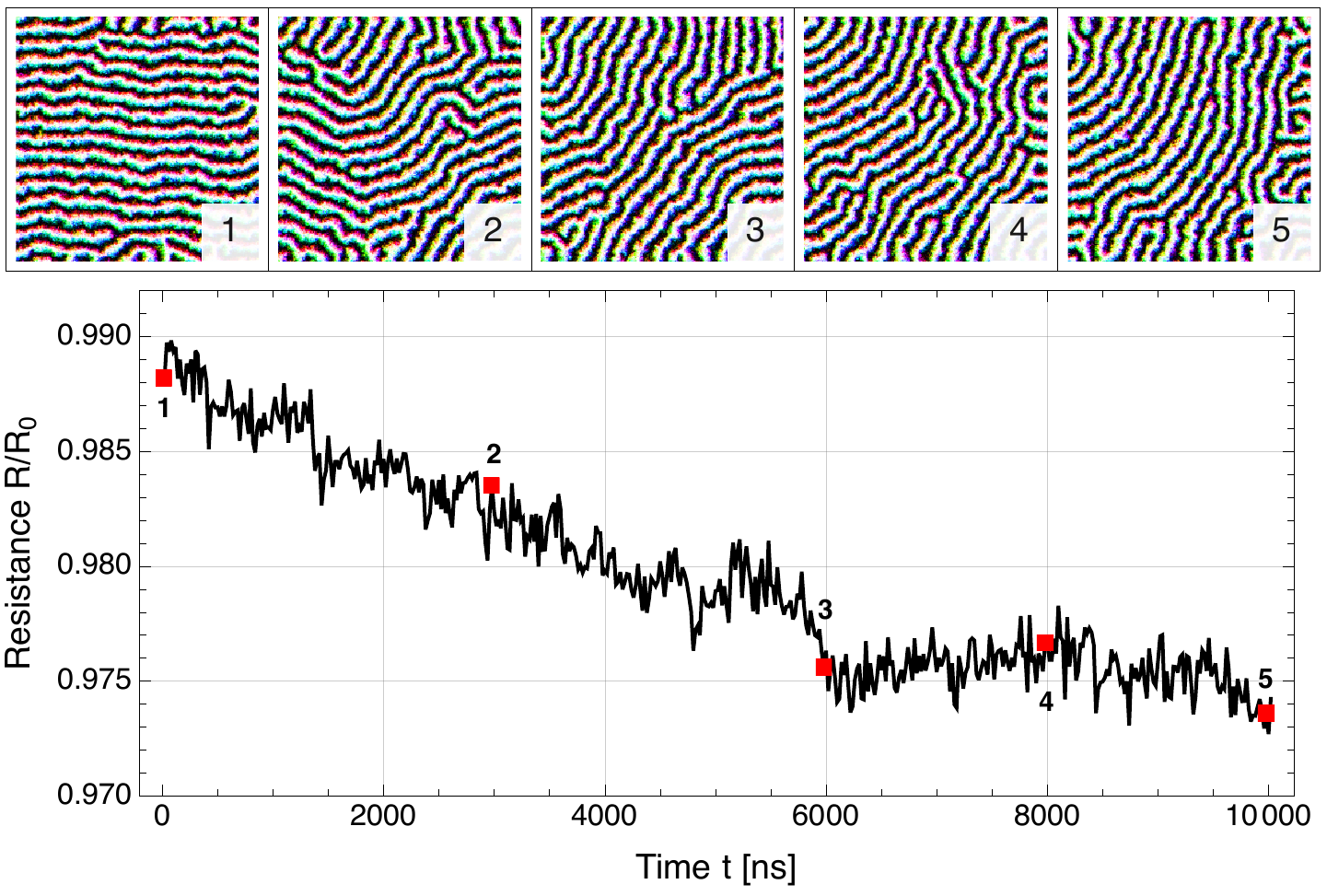}
	\caption{
        Fading memory in a large helitronic cell at finite temperature $T \approx T_c/2$.
        The upper panels show the magnetization corresponding to the states at the red squares in the graph.
        The graph shows the normalized resistance $R/R_0$ as function of time.
        The initial state, panel 1, is obtained after a current pulse with $\vec{j}\parallel-\bhat{x}$.
        Afterwards, the magnetization evolves at finite temperature without any applied currents.
        The system size is $16 \lambda \times 16 \lambda$.
        } 
	\label{fig6:temperature}
\end{figure}

We note that within our model the constant reset current $\vec{j}_\mathrm{r}$ is a simple solution for artificially fading memory.
However, by no means it is the energetically most efficient solution.
Instead, when a device is operated at finite temperature, the current-written long-ranged ordered states will naturally degrade over time due to thermal fluctuations.
We can then redefine the normal state of the device to be the thermodynamic ground state instead of the state with $\vec{q}\parallel\bhat{x}$.
The memory lifetime is then determined by the temperature $T/T_c$.
Also, the need for a constant reset current is eliminated, including the additional pair of electrical contacts.
Simulation results for the fading memory at finite temperature are shown in \fref{fig6:temperature}.
The temperature in the simulation is set to $T_\mathrm{sim}= 1290 \,\mathrm{K}$ and, for simplicity, we choose now $a_x = a_z = a_z = \lambda/16$.
The numerical lattice constant $a_\mathrm{sim}=\lambda/16\approx4.35\,\mathrm{nm}$ is much larger than the lattice constant of FeGe, $a_\mathrm{FeGe}=4.7\, \mathrm{\r{A}}$.~\cite{Richardson1967}
Hence, the simulated temperature effectively corresponds to $T=\frac{a_\mathrm{FeGe}}{a_\mathrm{sim}}T_\mathrm{sim}\approx 140\,\mathrm{K}$ \cite{Hahn2019} or, in units of the critical temperature $T_c=278.7\,\mathrm{K}$ \cite{Lebech1989}, $T/T_c\approx 0.5$.
At this temperature, an initialized state, panel 1, decays into the ground state, panel 5, on the time scale of $t_\mathrm{mem}\approx 6000 \,\mathrm{ns}$.
For long-term stable memristors, this time scale can be increased by lowering $T/T_c$.
On the contrary, it can be further decreased for fast neuromorphic applications by increasing $T/T_c$ which, however, also increases the noise of the resistance signal.
Note also, that in real systems with pinning the experimentally observed relaxation times are much larger.~\cite{Bauer2017,Dussaux2016}

\section{Conclusion}

We propose to exploit the orientation vector $\bhat{q}$ of a lamellar phase as a new order parameter in \emph{helitronic} devices.
Specifically, we study the helical phase in two-dimensional square-shaped samples of the chiral magnet FeGe as a conceptual model.
In micromagnetic simulations, we demonstrate the working principles of all-electrical binary and non-binary helitronic memory cells.
The state of the cells can be electronically read by measuring the resistance of the sample, exploiting that the resistance depends on the orientation of the helix due to AMR, see \sref{sec2}.
We propose four-terminal and two-terminal binary memory cells in \sref{sec3} and show that in FeGe they can be operated on the nanosecond timescale with current densities of the order of $10^{12}\sim10^{13}\,\mathrm{A/m^2}$ and cell sizes of $3\lambda \times 3\lambda$ where $\lambda\approx70\,\mathrm{nm}$ in FeGe.
Moreover, we propose non-volatile and artificial volatile non-binary memory cells with non-linear I-V-characteristics in \sref{sec4} as candidates for helitronic memristors or artificial neurons.
Our simulations for a cell as large as $16\lambda \times 16\lambda$ show that the operation with current densities of the order of $10^{11}\sim10^{12}\,\mathrm{A/m^2}$ is possible with current pulses on the nanosecond scale.

The prototypical chiral magnet FeGe on which we focus in this study should be regarded more as a convenient toy model than an actual high performance solution with practical applicability.
There is lots of room for improvement by optimizing the material, geometry, or physical effects exploited to switch and read the helitronic cells.
When reading the state of a cell electrically, the AMR should be maximized to ensure maximal fidelity. 
Weyl semimetals with large spin-momentum locking might hence be a good candidate.
A more precise differentiation between states might also be achieved by focusing on the two perpendicular orientations instead of ordered and disordered states.
For this, the recently discovered anisotropic DMI might be beneficial as it strongly pins the $\vec{q}$-vector of larger textures.~\cite{Karube2021}
In turn, extremely short ranged spirals experience strong anisotropic exchange which pins the $\vec{q}$-vector.
Smaller wavelengths allow for smaller system sizes.
Thus, smaller currents are needed to achieve a fixed switching time as the drift velocity of the texture is independent of its size.
Consequently, shorter wavelengths result in more dense information but also dissipate less energy.
Alternative  switching mechanisms, for example external fields, perpendicularly injected spin currents, spin-orbit torques, temperature gradients, anisotropy gradients, or electric fields in insulators, are also candidates for future investigations.
Yet, it should not be left unmentioned that pinning of the translational mode is usually large in lamellar phases.
Once it is under control, however, it helps to additionally increase the lifetime of the large-scale ordered states. 

In conclusion, we argue that the idea to use omnipresent lamellar phases to encode information may offer multiple benefits from the crossover between stray-field-free antiferromagnets and smoothly varying, well-studied ferromagnets.

\ack

N.T.B. and J.M. thank M. Garst for the fruitful discussions. N.T.B. particularly thanks P. Ge{\ss}ler for introducing him to MuMax3.


\section*{References}



\end{document}